\documentstyle[aps]{revtex}
\begin{document}
\draft
\twocolumn
\wideabs{
\title{On the existence of a Bose Metal at $T=0$}
\author{D. Das$^{1}$ and S. Doniach$^{1,2}$\\Departments of Applied Physics$^{1}$ and Physics$^{2}$\\
	Stanford University, Stanford, CA 94305. U.S.A.}
\date{\today}

\maketitle
\begin{abstract}
This paper aims to justify, at a microscopic level, the
existence of a two-dimensional Bose metal, i.e. a metallic phase made out
of Cooper pairs at $T=0$. To this end, we consider the physics of
quantum phase fluctuations in (granular) superconductors in the absence
of disorder and emphasise the role of two order parameters in the problem,
viz. phase order and charge order. We focus on the 2-d
Bose Hubbard model in the limit of {\em very large} fillings, i.e.
a 2-d array of Josephson junctions. We find that the algebra of phase
fluctuations is that of the Euclidean group $E_{2}$ in this limit, and 
show that the model is equivalent to
{\em two} coupled XY models in (2+1)-d, one corresponding
to the phase degrees of freedom, and the other the charge degrees of 
freedom. The Bose metal, then, is the phase in which both these
degrees of freedom are disordered(as a result of quantum frustration). 
We analyse the model in terms of its topological excitations and
suggest that there is a strong indication
that this state represents a surface of critical points, akin to the gapless
spin liquid states. We find a remarkable consistency of
this scenario with certain low-$T_{c}$ thin film experiments.
\end{abstract}
}
\narrowtext
\section{Introduction}

The superconductor-insulator(SI) transition in low-$T_{c}$ thin film 
systems \cite{gl} has drawn a lot of attention over the past couple of
decades. These systems
undergo transition from superconductor(SC) to insulator as the disorder, 
thickness or magnetic field is tuned. The problem has received a strong
impetus after the experiment by Goldman et. al.\cite{g2} on 
homogeneous lead and bismuth films, which went from a superconducting phase
to an insulating phase as a function of thickness, and which, right 
at the interface, was probably metallic.
Since electrons usually do not form a metallic state
in 2-d, it was argued that [Ref.3b] this $T=0$ transition is due to the
localisation of preformed Cooper pairs. It was also claimed that the
resistivity at the transition is universal[Ref.3b]. Similar SI transitions
have been observed in granular superconductors and Josephson Junction
arrays\cite{gl}. This scenario has been called into question after a 
recent magnetic field tuned experiment in the Mo-Ge sample\cite{kgb},
where the metal is no more a point in the phase diagram, but exists as 
a separate phase. We would like to point out that this is not the
first observation of a metallic phase in a two-dimensional, otherwise 
superconducting,
system. We came across, at least two separate instances of this
phenomenon in granular superconductors --- one in Ga films\cite{ag} and
the other in granular Pb films\cite{agb}, where a metallic phase is 
found to be sandwiched between the superconducting and the insulating 
phases. A similar observation has been reported in Josephson Junction
arrays\cite{jem}. Each of these observations has probably a detailed
explanation within the scope of the specific experimental system being 
measured. However,
there is something common in these systems which is quite hard to
overlook, viz. the preformed Cooper pairs are very much alive when the metallic 
phase is seen; and continue to exist till the insulating transition and beyond.
Motivated by this fact, we would like to ask the question --- is it 
possible for the charged bosons, i.e. Cooper pairs, to form an
incoherent metallic phase at $T=0$? The intriguing feature of this
phase is that although the bosons are mobile, they do not bose condense
at any temperature. Instead, they are dissipative and fail to drive a 
supercurrent even at $T=0$, unlike a superconductor. We regard this
phase as a Bose metal(BM). In this paper, we give arguments justifying the 
existence of a Bose metal in a physically realizable system.

The natural question, then is, what has been missing in the current theoretical
models where a metallic phase was not obtained. Our thought on this issue is the
following:
most of the earlier theories\cite{seb,mpa,mpf} tried to attack this problem
from the superconducting side of the phase diagram and projected it onto a 
basis diagonal in the phase states. In all these theories, there was a 
single order parameter, viz. $\psi=<e^{i\phi}>$, where $\phi$ refers to the
phase of the charge boson. In the superconducting phase, this order parameter
is well developed. At the SI transition, $\psi\rightarrow 0$ and the phases
of the charge bosons get scrambled. The scrambling of the phases\cite{seb,fg}
had so far been taken as the indication of onset of the insulating phase.
However, this is not enough to characterise the insulator. A Bose insulator(BI)
phase is characterised by an {\em extra order parameter}, viz. the charge 
density. It is like a charge density wave but built out of Cooper pairs.
This piece of physics has been missing in the existing
theories\cite{seb,mpa,mpf,fg}. 
The central point of this paper is that the phase fluctuation 
physics of superconductors should be viewed as a {\em two} order parameter
problem, viz. there is a competition between phase order and charge order.
It almost follows from this fact that the destruction of one order parameter
does not necessitate the growth of the other order parameter. This implies 
a possible existence of a disordered phase where both the order parameters are 
zero at $T=0$. We consider this to be a Bose metal phase.

Secondly, we work in the limit when the average filling of the bosons($n_{0}$)
per 
site is very large. In this limit, the Bose Hubbard model [Ref.3b], which 
plays a central role in the SI transition, becomes equivalent to a 
Josephson junction array (JJA) model. Most of the
existing work on the JJA model does not treat this large $n_{0}$
limit consistently.
In many cases, the constraint due to the average filling, eqn.(2) below
is neglected\cite{ottr};
in certain other cases, there is a tendency to replace the $E_{2}$ algebra,
which is the appropriate algebra for this limit (see sections II and V for
more details), by a qualitatively different algebra, viz.
$SU(2)$ algebra, which holds in the hard core limit, i.e. for small
fillings.
Hence, we shall give a quite detailed description of the basic formulation of
this limit in this paper. 

Thus, we consider a pure model of Josephson-coupled 
Cooper pairs with extremely large fillings, interacting via on-site and 
near neighbour repulsive forces in
two dimensions. This model captures the basic physics of granular
superconductors\cite{seb}, except that disorder is absent. It is applicable
to a limited extent to the Josephson junction arrays as well.
We are able to demonstrate that the physics of phase fluctuations,
within such a model, can be described by two coupled anisotropic XY models
in (2+1)-d, one corresponding to the charge degrees of freedom and the
other the phase degrees of freedom, where the coupling is ``XY-like''.
 The two
mechanisms which drive the two transitions are as follows -- disordering
of the vortices and the bose condensation of vortices.
Phase order is destroyed when the vortex-antivortex pairs unbind(or,
vortex loops blow up in (2+1)-d) as a result of the quantum fluctuations;
the charge order grows when the vortices Bose condense\cite{lee}.
In light of this, a search for a completely
disordered phase in the charge picture translates into a search for a
non superfluid(SF) liquid in the vortex picture. The presence of
unbound, as yet uncondensed, dissipative vortices makes the disordered
phase metallic. The two mechanisms mentioned above are separate processes
owing to the presence of retardation, or equivalently, dissipative
effects, coming from a gauge field mediated interaction in the vortex
picture (for more details see sections IId and IIIb). In the charge picture,
the existence of a disordered phase like that of a Bose
metal results from {\em quantum frustration} effects, i.e. the zero 
point motion is not strong enough to set up superconducting correlations
and the long range interactions are not sufficient to set up a charge ordered
state. This state seems to be intrinsically related to the 
(gapless) spin liquid states.
Although disorder is not explicitly present in our model, our results 
on the nature of the Bose metal phase can be readily generalised to the
case where disorder is present. We find a remarkable agreement of our 
predictions with the Gallium film experiment\cite{ag}. We believe this 
provides evidence of the existence of such a phase and supports 
the scenario
described above.  All of our considerations in this paper are restricted to
$T=0$ and zero magnetic field unless mentioned otherwise. This situation
is relevant to the experiments in Refs. \cite{ag} and \cite{agb}.

Before we go over to the main part of the paper, we would like to mention
that the idea of two order parameters is not completely new and has appeared
in the discussion of the hard-core limit of the Bose-Hubbard model\cite{gz}
and also in the gauge theory description of the Josephson Junction 
arrays\cite{it}.
The problem with the former is that superconductivity is never destroyed 
away from half-filling in their model (please see section IIb), while
the latter invokes the idea of self-duality
which implies logarithmic interaction among Cooper pairs and is not a good
representation of the realistic thin film samples.

Apart from the existence of a novel disordered Bose liquid state, there 
is another important feature which emerges from our work. It is well known 
that\cite{tv2} non-interacting electrons cannot support a metallic
state at low temperatures in 2-D in the absence of spin-orbit scattering. 
The situation is much less clear when both
interactions and disorder are present. This fact has been brought into
the limelight with the recent observation\cite{shr} of metal-like
states in two dimensional electron and hole systems in semiconductor based
materials. In the wake of these 
observations, our investigations suggest that a metallic state 
is a possibility if the electrons bind themselves into Cooper pairs
and behave as if they are bosons.

The plan of the paper is as follows. In section II, we introduce the
model and straighten out some of the basic issues which relate to the model.
Using a combination of qualitative and quantitative arguments, we explain why
the BM phase is feasible. The mapping onto the coupled XY models is
demonstrated here. We quantify the arguments in section III and
sketch the phase diagram of the model. Section IV contains
an estimate of the resistivity of the Bose metal phase and a
comparison with the Gallium film experiment. The thermodynamics of 
this strange metallic state and its connection with the spin liquids
are discussed in section V. We wind up with a discussion of
certain relevant issues and conclusions in section VI. Some of the
calculational details may be found in the appendices.

\section{The foundations}
\subsection{The model}

We shall consider the following model Hamiltonian in this paper ---
\begin{eqnarray}
H&=&-J\sum_{<i\alpha>}cos(\phi_{i}-\phi_{i+\alpha}) + 
V_{0}\sum_{i}(\hat{\delta n_{i}})^{2} \nonumber \\ &+&
V_{1}\sum_{<i\alpha>}(\hat{\delta n_{i}} +
\hat{\delta n}_{i+\alpha})^{2} - 
\bar{\mu}\sum_{i}\hat{\delta n_{i}} - \mu N n_{0}
\end{eqnarray}
where $\hat{\delta n_{i}}=\hat{n_{i}}-n_{0}$, with $\hat{n_{i}}=$ number 
density operator, $n_{0}=$ neutralising background charge density(or, 
equivalently, the average density of Cooper pairs), $N=$
number of lattice sites, and $\bar{\mu}=\mu+V_{0}$ the renormalised chemical
potential in the problem. Also, $\alpha = \hat{x},\hat{y}$ refers to the
spatial unit vectors. The first term in eqn.(1) represents the kinetic
energy of the bosons(Cooper pairs for our case), the second term the onsite
repulsion among them(which should be non-zero to prevent any collapse of
the bosons onto a single site), the next term the repulsion among the
nearest neighbours which acts to set up a charge order in the system, 
and the last two terms are the chemical potential terms.
Eqn.(1) has to be supplemented with the constraint $\frac{1}{N}\sum_{i}
<\hat{n_{i}}> = n_{0}$, which in terms of the charge fluctuation operators 
convert into 
\begin{equation}
\sum_{i}<\hat{\delta n_{i}}> = 0
\end{equation}
This equation implies $\bar{\mu}=0$ and, in the rest of the paper, we shall
forget about the chemical potential terms completely.

On the experimental side, the three parameters of the Hamiltonian (1) can be
determined in the following way\cite{es}.
\begin{equation}
   J = (R_{Q}/2R_{n}) \Delta_{0} 
\end{equation}
Here $R_{Q}=(h/4e^{2})=6.45K\Omega$, $R_{n}=$ normal resistance of the film,
and $\Delta_{0}=$ pairing gap. The interaction constants $V_{0}$ and
$V_{1}$ are related to the inverse of the capacitance matrix $C_{ij}$ of 
the grains.

\subsection{The commutation relations}

The model discussed in eqn(1) needs to be supplemented with 
the phase fluctuation algebra, which constitutes the appropriate commutation 
relations for this problem:
\begin{equation}
[\hat{\delta n_{i}},\hat{\phi_{j}}] = i\delta_{ij}
\end{equation}
Eqn.(4) implies an angular momentum representation[Ref.3a]
\[ \hat{\delta n_{i}} = i\frac{\partial}{\partial\phi_{i}}. \]
Further, let us define the operators ---
\begin{equation}
 L = i\frac{\partial}{\partial\phi} \textup{, and }
   P = e^{i\phi}. 
\end{equation}
Thus, eqn.(4) can be recast as ---
\[ [L,P] = -P  \makebox[1in][r]{(6a)}   \]
\[ [L,P^{\dagger}] = P^{\dagger} \makebox[1in][r]{(6b)}\]
\[ [P,P^{\dagger}] \simeq 0  \makebox[1in][r]{(6c)} \]
with $PP^{\dagger}=I$. It is well known in quantum optics that the phase
operators being ladder operators[as is seen from eqns.(6a) and (6b)] 
usually do not commute\cite{opt}. However, in the large 
$n_{0}-$limit, they do (see appendix F). Thus, eqn.(6c) is strictly valid 
in the large $n_{0}-$limit
of the problem and, hence, our discussion holds good in this limit
only. Now, eqns.(6) constitute the 
{\em algebra of the Euclidean group} $E_{2}$\cite{e2},
the group of translations and rotations in 2-d, with the square
of linear momentum($P$) restricted to unity for our case. It deserves to be
mentioned here that the $SU(2)$ algebra used in the context of the hard-core
limit of the Bose problem\cite{gz} is distinctly different from this algebra 
and is obtained in the opposite limit, i.e. the small $n_{0}-$limit of the
Bose Hubbard model. The effects of the change in the group structure are
quite significant. In the hardcore model, away from half filling, 
increased interactions changes the SF to a supersolid phase. Thus, 
superfluidity is never destroyed away from half filling. 
On the other hand, our model allows superfluidity to be
quenched at arbitrary fillings. Further, the conserved quantities supported
by the two algebras are different: the invariant of $E_{2}$ algebra is
(square of linear momentum) $P_{x}^{2}+P_{y}^{2} = p^{2}$ ($=1$ for our 
case), as compared to $S_{x}^{2}+S_{y}^{2}+S_{z}^{2} = s(s+1)$ of the $SU(2)$
algebra. Thus, unlike the $SU(2)$ case, (a) here the constraint on the 
z component of the spin is much weaker and (b) the z component of the spin
enters very anisotropically in the algebra compared to the x and y components
of the spin. (Here operator $L$ is referred to as the z component of spin for
$E_{2}$ algebra; see section V for more details.) Hence, in the
disordered/symmetric phase, although the x and y components of $E_{2}$ spin
might acquire a gap like the $SU(2)$ spin, the z component of the former 
might remain gapless unlike the latter.
Put in simple words,
there may be a length scale determining the local superfluidity
in this phase and as yet no length scale associated with the
charge ordering in the system. A calculation of the charge charge 
correlation function indeed justifies this, as discussed in section V.
Thus, given a disordered phase within a model built out of
$E_{2}$ spin operators, it can support gapless excitations.
We shall
give detailed arguments in sections IID and III as to why the model (1)
along with commutation rules (6) contain a completely disordered phase.

Now, if we rotate the charge fluctuations $L_{i}$ at each site, viz.
\[L_{i} \rightarrow e^{i\vec{Q}.\vec{r_{i}}}L_{i}, \textup{ where }
\vec{Q}=(\pi,\pi)\]
we obtain the Hamiltonian as
\setcounter{equation}{6}
\begin{equation}
 H = \sum_{k}J_{k}P_{k}^{\dagger}P_{k} + \sum_{k}V_{k}L_{k}^{\dagger}L_{k},
\end{equation}
where
\[ J_{k} = -J(cosk_{x} + cosk_{y}) \]
\[ V_{k} = (V_{0}+4V_{1})- 2V_{1}(cosk_{x} + cosk_{y}). \]
and $P, L$ couple through the commutation relations (6).
In this form, it assumes the shape of a 2 order parameter problem.(Macroscopic
occupation of $k=0$ mode of $P$ reflects superconductivity
 and that of $L$ a charge density wave.) However, because of commutation 
rules (6), it is very hard to 
diagonalise this hamiltonian in this form. So, we seek alternative means.

\subsection{The charge picture}

In this section, we demonstrate that the model Hamiltonian (1) is equivalent 
to two coupled XY models in (2+1)-d. To do this, we first write the model (1)
in terms of a path integral representation\cite{e2,seb2} 
keeping the commutation 
relations (6) in mind. The partition function $Z = Tre^{-\beta H}$, where 
$\beta$ is the inverse temperature ($\hbar$ and $k_{B}$ are taken to be one,
unless mentioned otherwise), can be written as ---
\[Z = \sum_{\{m_{i}(\tau)\}}\int_{0}^{2\pi}{\mathcal D}\phi_{i}(\tau) e^{-S},\]
where
\begin{eqnarray}
S&=&i\int_{0}^{\beta}\sum_{i}m_{i}(\tau)\frac{\partial \phi_{i}}{\partial\tau}
+ \int_{0}^{\beta}d\tau[-J\sum_{i\alpha}cos(\phi_{i}(\tau)-\phi_{i+\alpha}
(\tau)) \nonumber \\
&+& V_{0}\sum_{i}m_{i}^{2}(\tau) + V_{1}\sum_{i\alpha}
(m_{i}(\tau)+m_{i+\alpha}(\tau))^{2}]
\end{eqnarray}
with periodic boundary conditions in the imaginary time direction implied. 
Here $\alpha = \hat{x},\hat{y}$ refers to the nearest neighbours in the
space direction and $m_{i}(\tau)$ are integers meaning the change in the
number of Cooper pairs from the average at the site $i$. 
We shall consider only the case 
$T=0$, so that the integral over imaginary time extends to infinity. Next,
we discretize the time axis with an interval $\Delta\tau$ and rescale
the lattice constant in the space directions to be unity. Further, we rotate
the integers $m_{i}(\tau) \rightarrow e^{-iQ.r_{i}}m_{i}(\tau)$, where
$\vec{Q}=(\pi,\pi)$. Thus, we obtain ---
\begin{eqnarray}
S&=&i\sum_{i}e^{-iQ.r_{i}}m_{i}(\nabla_{\tau}\phi_{i}) + V_{0}\Delta\tau
\sum_{i}m_{i}^{2} \nonumber \\
&+&V_{1}\Delta\tau\sum_{i\alpha}(\nabla_{\alpha}
m_{i})^{2} - J\Delta\tau\sum_{i\alpha}cos(\nabla_{\alpha}\phi_{i}).
\end{eqnarray}
Here summation over $i$ refers to the time axis as well, and so does the 
index $i$ in $m_{i}$ and $\phi_{i}$. The derivatives $\nabla_{\mu}$ in eqn.(9)
and also in what follows are lattice derivatives\cite{ld}. To avoid
any confusion, we shall reserve the notation $r_{i}$ for the spatial
coordinates of the $i^{th}$ point and $x_{i}$ its space-time coordinates.

To show that the action (9) is equivalent to 2 coupled XY models, we follow
the following sequence of steps. (1)We decouple the $(\nabla_{\alpha}m_{i})^{2}$
term using a Hubbard-Stratanovich field $p_{i\alpha}$, viz.
\[ \int{\mathcal D}p_{i\alpha} e^{-(1/4V_{1}\Delta\tau)\sum_{i\alpha}
p_{i\alpha}^{2} - i\sum_{i}m_{i}(\overline{\nabla}_{\alpha}p_{i\alpha})}\]
(2)First, we notice that the coupling term with $m_{i}$ is invariant under
shifts of $p_{i\alpha}$ by $2\pi$. So, we break up the intgral over
$p_{i\alpha}$ from $-\infty$ to $+\infty$ into that of periods of $2\pi$.
Further, we split up $p_{i\alpha}$ into a curl and a gradient part. Since
the divergence of $p_{i\alpha}$ couples to $m_{i}$, only the gradient part
enters the dynamics. Thus, we obtain for this part,
\begin{eqnarray*}
 \sum_{\{l_{i\alpha}\}} &\int_{0}^{2\pi}& {\mathcal D}\theta_{i} 
e^{-i\sum_{i}m_{i}\nabla^{2}\theta_{i} - (1/4V_{1}\Delta\tau)
\sum_{i\alpha}(\nabla_{\alpha}\theta_{i}-2\pi l_{i\alpha})^{2}} \\
& \simeq &
\int_{0}^{2\pi}{\mathcal D}\theta_{i} e^{(1/2V_{1}\Delta\tau)
\sum_{i\alpha}cos(\nabla_{\alpha}\theta_{i})}
e^{-i\sum_{i}m_{i}\nabla^{2}\theta_{i}}, 
\end{eqnarray*}
where we have used an (inverse) Villain transformation\cite{it}. (3) Now, using
an (inverse) Villain transformation again, one can execute the sum over 
integers $m_{i}$ ---
\begin{eqnarray*}
 \sum_{\{m_{i}\}}&exp&(-i\sum_{i}m_{i}(e^{-iQ.r_{i}}\nabla_{\tau}\phi_{i}
+ \nabla^{2}\theta_{i}) - V_{0}\Delta\tau\sum_{i}m_{i}^{2}) \\ &\simeq&
e^{(1/2V_{0}\Delta\tau)cos(e^{-iQ.r_{i}}\nabla_{\tau}\phi_{i} + 
\nabla^{2}\theta_{i})}  
\end{eqnarray*}
Putting all these together, one obtains ---
\[ Z = \int_{0}^{2\pi}{\mathcal D}\phi_{i}{\mathcal D}\theta_{i} e^{S}  \]
with
\begin{eqnarray}
S&=&J\Delta\tau\sum_{i\alpha}cos(\nabla_{\alpha}\phi_{i}) \nonumber \\
&+& \frac{1}{2V_{0}
\Delta\tau}\sum_{i}cos(e^{-iQ.r_{i}}\nabla_{\tau}\phi_{i} + \nabla^{2}
\theta_{i}) \nonumber \\ &+&\frac{1}{2V_{1}\Delta\tau}\sum_{i\alpha}
cos(\nabla_{\alpha}\theta_{i})
\end{eqnarray}
where the phase $\phi$ is associated with superfluidity and the phase 
$\theta$ with charge density wave.
Eqn.(10)  explicitly shows that the phase fluctuation physics of (granular) 
superconductors, in the absence of disorder, is equivalent to
two coupled XY models in (2+1)-d, where the coupling
is XY-like. A few comments are in order.
In the limit $V_{1}=0$, we have $\theta_{i}=0$, the $\theta$-terms
in action (10) drop out, and we obtain a single XY model in (2+1)-d, as
has been discussed previously\cite{seb,fg,seb2}. Also, the coupling term (the 
second term) is highly anistropic in $\phi$ and $\theta$, explicitly 
breaking self-duality in this system, {\em assumed} in Ref.\cite{it}. 
Eqn.(10) is one of the key results of this
paper. Thus, we see that the destruction of superfluid state is driven
by one XY model, whereas the other XY model characterises the growth of 
the charge ordered state. And, hence, as the parameters are tuned one 
transition does not necessarily accompany the other.

\subsection{The vortex picture}

Although the action (10) shows that the superfluid order and charge order 
are driven by different XY models, it is not clear whether the coupling
between them, which is quite complicated, guarantees a completely disordered
phase. In order to answer such a question, 
we now consider the model (1) in the dual picture, i.e. of the vortices. To
do this, we invoke a duality transformation\cite{seb,it,lee,seb2}.
Starting from action(9)\cite{n1}, one can show that
\begin{eqnarray}
S&=&\frac{\pi^{2}}{ln(\frac{2}{J\Delta\tau})}
\sum_{q,\omega}j^{0}_{q,\omega}\frac{1}{\hat{q}^{2}}j^{0}_{-q,-\omega}
\nonumber \\ &+&
\frac{\pi^{2}}{ln(\frac{2}{J\Delta\tau})}
\sum_{q,\omega}j^{\alpha}_{q,\omega}G(\omega,q)j^{\alpha}_{-q,-\omega},
\end{eqnarray}
where $j^{\mu}_{q,\omega}$ are Fourier transforms of the integer vortex variables
$j^{\mu}_{i} (j^{0}_{i}=$ vortex density, $j^{\alpha}_{i}=$ vortex current),
and
\begin{equation}
G^{-1}(\omega,q) = [\hat{\omega}^{2}+\frac{V_{0}\Delta\tau}{ln(\frac{2}{J\Delta\tau})}\hat{q}^{2}
+ \frac{V_{1}\Delta\tau}{ln(\frac{2}{J\Delta\tau})}\hat{q}^{2}(\hat{q}-Q)^{2}]
\end{equation}
The intervening steps are quite standard\cite{sav} and are discussed 
in Appendix A. $\hat{\omega}, \hat{q}$ refer to lattice frequency and momentum
respectively[Appendix A]. One may like to note here that there is no Magnus
force term on the vortices, corresponding to the disappearance of $n_{0}$
from the problem in the large $n_{0}$ limit.

In the previous subsection, we observed that when $V_{1}\rightarrow0$, the
model is equivalent to a single XY model rather than two coupled XY models.
The question, obviously is, how this change is reflected
in the dual picture. The basic answer lies in the appearance or disappearance
of the zero point motion term for the vortices(more popularly known as the
vortex mass term). To show this, we look at the long wavelength low
frequency modes. In 
the limit $\omega,q\rightarrow0$ and $\omega\ll c_{s}q$(where $c_{s}$ is
the plasmon velocity), the Green's function $G(\omega,q)$ 
splits up into 
a singular part $G_{s}(\omega,q)$
and a constant part $G_{0}$, viz.
\begin{equation}
G(\omega,q) \simeq G_{s}(\omega,q) + G_{0},
\end{equation}
where
\[ G_{s}(\omega,q) = 1/(\omega^{2}+c_{s}^{2}q^{2}), \textup{ }
   G_{0} = b^{2}/8(b^{2}+c^{2})^{2}   \]
with 
\[ c_{s}^{2} = b^{2} + c^{2}, \]
\[ b^{2} = 8 \frac{V_{1}\Delta\tau}{ln(\frac{2}{J\Delta\tau})} \textup{, }
   c^{2} = \frac{V_{0}\Delta\tau}{ln(\frac{2}{J\Delta\tau})} \]
Thus the action is 
\begin{eqnarray}
S&=&\frac{\pi^{2}}{ln(\frac{2}{J\Delta\tau})}
\sum_{q,\omega}j^{0}_{q,\omega}\frac{1}{q^{2}}j^{0}_{-q,-\omega} \nonumber \\
& + & \frac{\pi^{2}}{ln(\frac{2}{J\Delta\tau})}
\sum_{q,\omega}j^{\alpha}_{q,\omega}[G_{s}(\omega,q)+G_{0}]j^{\alpha}_{-q,-\omega},
\end{eqnarray}
Since $G_{0}$ is a constant,
 it allows us to identify this term as the vortex kinetic energy,
i.e. mass term. From (14) and (13), we notice that this term does not 
exist when $V_{1}=0$. Thus, we have shown that the change of the 
nature of the XY models is tied to the existence or non-existence of a
vortex mass term.

We now look at the other terms in the action (14). 
The first term in the action is the 
usual logarithmic interaction term among the vortices, and the second term,
because of the retardation effects, leads to a dissipative term for the
vortices\cite{es}. In the limit $\mid r_{i} - r_{j}\mid \ll c_{s}\mid\tau -
\tau'\mid$(which is the same quasistatic limit discussed so far), this 
part of the action takes the form ---
\[ \frac{\pi^{2}}{c_{s}^{2}ln(2/J\Delta\tau)}\sum_{i,j}q_{i}q_{j}
\sum_{\omega} \omega^{2}ln(\frac{1}{\mid\omega\mid}) r_{i,\omega}
r_{j,-\omega}  \]
where $q_{i}=\pm 1$ refers to the charge on the $i^{th}$ vortex.
Physically, the aforesaid limit corresponds to the slow motion of vortices.
The source of this heat bath(or, dissipation) is the gauge field
(discussed in Appendix A), or more precisely, the transverse modes 
arising from the quantum fluctuations in the system. (There will be
additional retardation effects in a real system from external 
heat bath mechanisms, e.g. electrons in the vortex core, etc.) These
features of vortices in granular superconductors have been discussed 
previously by Eckern and Schmid\cite{es}.

One may also undo some of the steps in Appendix A, and recast eqn.(14) in 
terms of a gauge field with appropriate action as ---
\begin{eqnarray}
S&=&ln(\frac{2}{J\Delta\tau})\sum_{q,\omega}[q^{2}A^{0}_{q,
\omega}A^{0}_{-q,-\omega} + (\omega^{2} + c_{s}^{2}q^{2})A^{\alpha}_{q,
\omega}A^{\alpha}_{-q,-\omega}] \nonumber \\ 
&+& 2\pi i\sum_{i}j^{\mu}_{i}A^{\mu}_{i} + 
\pi^{2}\frac{G_{0}}{ln(\frac{2}{J\Delta\tau})}
\sum_{q,\omega}j^{\alpha}_{q,\omega}j^{\alpha}_{-q,-\omega} 
\end{eqnarray}
(with the gauge condition $\nabla_{\alpha}A^{\alpha}_{i} = 0$ imposed).
A few comments are in order. Eqn. (15) shows that,
considered in the vortex picture, the quantum phase fluctuations in a 
2-d superconductor, as described by model (1), is equivalent to
a {\em two}-component {\em quantum} plasma (bosons of two flavours, 
viz. vortices and antivortices) moving in a
fluctuating gauge field $A^{\mu}$\cite{il}. Secondly,
the above scenario is a simple quantum mechanical extension of classical
phase fluctuations in a 2-d superconductor, which is described by 
a two component {\em classical} plasma 
undergoing screening by a static electric field ($\vec{E} = -\vec{\nabla}
A_{0}$), as described by Kosterlitz and Thouless(KT)\cite{kt}. The effect
of including quantum mechanics in the problem, apart from bringing up
the importance of the quantum statistics of vortices, is to make the
electric field dynamical, viz. $\vec{E} = - \vec{\nabla}A_{0} - (1/c_{s})
\partial\vec{A}/\partial\tau$ with a magnetic field
$\vec{B} = \vec{\nabla}\times\vec{A}$. Whereas the importance of the
statistics is to allow for the superfluidity of the vortices, an 
important consequence of the dynamical nature of the electromagnetic 
field is that there are retardation effects, viz.
\[ \sum_{i,j}\int d\tau d\tau'\frac{\dot{\vec{r}}_{i}(\tau).
 \dot{\vec{r}}_{j}(\tau')}
  {\sqrt{(\vec{r}_{i}(\tau)-\vec{r}_{j}(\tau'))^{2} + c_{s}^{2}(\tau 
 - \tau')^{2}}}, \]
which break Galilean invariance. (This feature is intensified in a real 
system by external heat bath mechanisms mentioned before\cite{es}.)
This is not surprising because the
action (15) has the structure of Maxwell's action, which is reputed to
have Lorentz invariance but lacks Galilean invariance. Now the
absence of Galilean invariance will have a strong effect on our 
system, because it is bosonic. It is well known that {\em all} the
delocalised bosons will condense into the superfluid state only if 
the system is Galilean invariant\cite{pns}. If this invariance is
absent in a Bose system, then as the relevant parameter is tuned, the
(vortex) condensate is gradually depleted and at one point the 
superfluidity will be completely lost.  
At this stage, it is important to recollect what 
important processes are going on in this system. There are two of them:
destruction of vortex (and antivortex) superfluidity (owing to retardation
effects) and the binding of
vortex-antivortex pairs, corresponding respectively to the destruction of
charge order and the growth of phase order. Now, these two processes
are controlled effectively by two separate parameters, viz. $c^{2}/
g^{2} \sim (V_{0}+8V_{1})\Delta\tau$ (corresponding to the strength of
the retardation effects) and $g^{2} \sim J\Delta\tau$ (corresponding to 
the strength of the logarithmic interaction)
respectively, where $\Delta\tau \sim 1/\Delta_{0}$\cite{bd} (please
refer to appendix D and section III for the notation and appropriate
details). As a result, the vortices(and antivortices) do not necessarily
condense into the superfluid state as soon as they unbind.
 This leads to the possibility of a non-SF
vortex liquid, or equivalently, a BM phase. In fact, that's what we find 
when we quantitatively evaluate these processes
in section III. There is a simpler way to see 
what is happening here. As we noted earlier, following eqn.(14), 
the kinetic energy
of vortices originates from two sources --- (a)quantum zero point motion
(the $G_{0}$ term) and (b)action of an {\em effective}
heat bath (the $G_{s}$ term). In the delocalised state,
if the source (a) dominates, the vortices move coherently, and since they
are bosons, they form a superfluid. On the other hand, when the source
(b) dominates, because of the random nature of the effective
heat bath, the motion
of the vortices is necessarily incoherent, and one is in a metallic phase.
In this phase, no charge order is set up and one obtains a Bose metal phase.

These features of dual vortices are not special to a lattice model, but
observable in the continuum formulation as well, as discussed in
Appendix B.

\section{The Phase diagram}

In the previous section, we argued why the Hamiltonian described by eqn.(1)
may contain an incoherent metallic phase. In this section, we quantify
these arguments by calculating the phase diagram of model (1):
we shall locate the phase boundary where 
superconductivity is destroyed and the one where charge order is established.

\subsection{Destruction of charge superfluidity}

This is done non-perturbatively by estimating where the vortex loops 
blow up in (2+1)-d\cite{n2}. This happens when the entropy of the loops
overcomes their interaction energy\cite{n3}. A good estimate of the 
interaction energy is the self energy of the loops, simply
because dipole-dipole interactions fall off as $1/r^{3}$ and the mutual
interaction energy of the links in a loop is much smaller than the
self-energy of the loops when the loops are fairly large. Thus, the 
effective free energy of the loops is given by
\begin{equation}
{\mathcal F} = [\frac{\pi^{2}}{ln(\frac{2}{J\Delta\tau})}G(0) - \mu_{l}]N
\end{equation}
where $N$ is the number of the links in a vortex loop, $G(0)=$ diagonal
part of the Green's function $=\int_{-\pi}^{\pi}\frac{d\omega}{2\pi}
\frac{d^{2}q}{(2\pi)^{2}}G(\omega,q)$, obtained from eqn.(12),
and $\mu_{l}=$ entropy of the loops
$= ln3$ for our case\cite{n4}. An estimate of $G(0)$ is given in the
Appendix C. The loops blow up when ${\mathcal F} < 0$. Thus,
superconductivity is destroyed when\cite{n5}
\begin{equation}
(V_{0}/J) + 8(V_{1}/J) > \tilde{b_{0}},
\end{equation}
where $\tilde{b_{0}}=2(b_{0}/\mu_{l})^{2}$ and
$b_{0}=$ a number of order unity $\simeq$ 3.1725, defined by eqn.
(C6) in appendix C. This phase
boundary has the character of (2+1)-d XY model, at least when 
$V_{1}\ll V_{0}$ (as seen from eqn.(10)), and hence, the superfluid density
changes continuously across this transition line.

\subsection{Destruction of vortex superfluidity} 

It is well established that the existence of charge order implies 
superfluidity of vortices and vice versa\cite{lee}. We shall follow this
notion here and estimate the growth of charge order in terms of the
superfluidity of vortices. We mentioned in section IId that
the vortices and antivortices move in the presence of a {\em dynamical}
gauge field and argued that because of the latter there are 
retardation effects which deplete the (vortex) Bose condensate (due to
broken Galilean invariance).
As a result, the vortices and antivortices do not necessarily
condense into a superfluid state as soon as they unbind. This physics of
suppression of Bose condensation as a result of gauge field
fluctuations is not new but has been explored substantially in the 
context of spin charge separation theories in high $T_{c}$
superconductors\cite{rvb,il}. The discussion here is very similar in
spirit to that piece of work.

To estimate the strength of the parameters where vortex superfluidity is
destroyed, we follow the self-consistent functional approach of Ioffe 
et. al.\cite{il}. They did the calculation for a very similar 
piece of physics,i.e. how the coupling to a gauge field can kill
superfluidity in a bosonic system (with logarithmic interaction) as a
result of broken Galilean invariance stemming from retardation effects.
We refer the reader to that paper for a full description of this technique.
A short discussion of this is given in Appendix D. We can follow 
their approach here, simply because the lattice action and the
continuum action have identical structure in the
low frequency long wavelength limit, as discussed in Appendix B.
To do this, we
first replace the two component plasma by a one component plasma, i.e.
charges(vortices) of one flavour moving in the background of fixed 
neutralising charges of the other flavour. This approximation is very 
standard and captures the salient features of the problem, until and 
unless the plasma is extremely dense\cite{si}. Then, the results of sec.V 
in Ref.\cite{il} can be directly carried over here with the identification
of the Coulomb interaction parameter $\alpha_{c}$ and transverse gauge
field coupling constant $\alpha_{g}$ as\cite{n5} (Appendix D) ---
\begin{eqnarray}
\alpha_{c} &=& (\pi^{2}/2)v_{1}/(v_{0}+8v_{1})^{2}  \nonumber \\
\alpha_{g} &=& (1/\pi n_{v})[1+(v_{0}/8v_{1})]  
\end{eqnarray}
where $n_{v}=$ average vortex density and 
$v_{i} = (V_{i}/J)$,$(i=0,1)$. Since what counts in the destruction of
vortex superfluidity are the free vortices(and antivortices), we take 
$n_{v} \sim n_{f}$, which goes inversely as the square of the correlation
length $\xi_{+}$ which diverges at the SC-BM boundary from the BM side.
Thus, from (17), $n_{f} \sim [\bar{b}_{0}(V_{0}+8V_{1})/J - 1]^{2\nu}$,
($\bar{b}_{0} = 1/\tilde{b}_{0}$)
with $\nu \sim 2/3$, since this phase boundary has the character of (2+1)-d
XY model. Also, if we are above the phase boundary of eqn.(17) (see fig.1),
we have $\alpha \ll 1$ as in Ref.\cite{il}, where $\alpha = \sqrt{\alpha_{c}}$.
Since as we shall note below that the transition from SF to non-SF
state takes place at a small value of $\alpha$, this assumption of small
$\alpha$ is self-consistent.

Let us first consider the simple case when we are slightly above the phase 
boundary (16) and along the $V_{1}$ axis (see fig.1), so that $V_{0} = 0, 
V_{1} \sim
2.1J$. Then, we have, $\alpha \simeq 0.2$ and $\alpha_{g}\gg 1$; 
the calculation of Ref.\cite{il}, in that case, suggests we are in the
disordered BM phase. Thus, there is at least a small region close to the
phase boundary (17), where the system is metallic.

We now complete the calculation of phase boundary of BM-BI transition,
following Ref.\cite{il}. Since $0 < n_{v} < 1$, from (18), $\alpha_{g} > 0.32$.
From fig.7 of Ref.\cite{il}, one can see that in this regime, the 
phase boundary 
tends to saturate at $\alpha = \alpha_{cr} \simeq 0.08$. Actually, 
for larger values 
of $\alpha_{g}$, $\alpha_{cr}$ is a little less; but, the point is that
$\alpha_{cr}$ is always finite, however small it be, for large values of
$\alpha_{g}$. That is, the vortices always form a superfluid for small enough
$\alpha$. A physical way of seeing why there is always superfluidity for 
small enough Coulomb repulsion is as follows. The strength of the
Coulomb repulsion is controlled by the effective charge $g$ (see appendix
D), which also
controls the strength of the coupling between the particles(vortices) and the
transverse gauge field. So, when the magnitude of Coulomb repulsion is
small, $g$ is small, and hence, the coupling with the transverse gauge field
is small as well. As a result, the particles do not feel the effect of
retardation strongly enough in this limit and condense into a superfluid
phase. Thus, the vortices are in a superfluid state when
$\alpha < \alpha_{cr}$, which means ---
\begin{equation}
\frac{1}{(8V_{1}/J)}[(V_{0}/J) + (8V_{1}/J)]^{2} > c_{0}
\end{equation}
where $c_{0} = 4(\pi/8\alpha_{cr})^{2} \simeq 96$ for $\alpha_{cr} 
\simeq 0.08$. This calculation implies a jump in the (vortex)
superfluid density at this phase boundary\cite{il}. This means that the phase 
transition
is either of first order or has a KT character. More calculations are 
necessary to resolve this point.

We display a schematic phase diagram determined by the eqns.(17) and (19)
in fig.1. Eqn.(17) is represented by the curve LXM and (19) by ZYX. They
seem to meet at a {\em tricritical} point X. Since the retardation effects are
very strong in this model (meaning that the value of $\alpha_{cr}$ is small), 
the constant $c_{0}$ in eqn.(19) is roughly
an order of magnitude larger than the constant $\tilde{b_{0}}$ appearing 
in eqn.(17); and hence we expect that this crossing between the two curves
will always occur, implying the existence of
the Bose metal (see section VI as well). The various phases
determined by these eqns. are as marked in fig.1 --- when the zero
point motion of the Cooper pairs is large, the system is superconducting
(region LXMO); when the interactions dominate, the system is charge
ordered and insulating as a result --- we call it a Bose insulator(BI);
and in the intermediate region, the system is disordered and, hence,
metallic (please see section IV below) --- the BM phase (region ZYXL).

What one sees on the phase diagram of fig.1 is that the metallic phase is
more prominent towards the $V_{1}$ axis rather than the $V_{0}$ axis.
Further, from eqn.(1) we note that the $V_{1}$-term contains both 
on-site and nearest neighbour repulsion energies. This shows that the 
Bose metallic phase is to be expected in cases where these energy scales
are of {\em comparable} order of magnitude, a situation well represented by
the granular superconductors\cite{jem}.

We can provide a physical explanation in the charge
picture as to why the metallic phase opens up along the $V_{1}$ axis:
let us focus on eqn.(1) and say that superconductivity is already 
quenched. First consider the case $V_{0}$ large and $V_{1}$ small, so
that we are along (or, close to) the $V_{0}$ axis. Naively, one
would expect from the $V_{1}$ term that $\delta n_{i} + \delta n_{i+
\alpha} \sim$ large. But, this costs a large energy from the $V_{0}$
term. As a result, what is favoured is $\delta n_{i} \simeq 0$, i.e. 
$n_{i}\simeq n_{0}$, a (rather trivial) charge ordered state. Now, consider
the opposite case: $V_{1}$ fairly large and $V_{0}$ small, so that we
are close to (or, along) the $V_{1}$-axis. Since $V_{1}$ is appreciable,
eqn.(1) suggests $\delta n_{i} + \delta n_{i+\alpha}\simeq 0$. A non-trivial
configuration may be $\delta n_{i} = -\delta n_{i+\alpha} = 1$, for 
some $i$'s and zero otherwise. This state represents an RVB-like 
state of fluctuating
charge (particle-hole) dipoles, the equivalent of spin singlets here
(please see section V for more on this point). $V_{0}$ being small, the
$V_{0}\sum_{i} (\delta n_{i})^{2}$ term does not cost much energy for
this kind of state. This is the disordered BM phase. When $V_{1}$ is
increased further, the dipoles freeze into a charge ordered solid. Thus,
very crudely speaking, the smallness of $V_{0}$ is a source of frustration
in this model for finite $V_{1}$.

Since $J$ is
inversely proportional to the normal resistance $R_{n}$ [eqn.(3)], as
the normal resistance of a thin film is tuned in an experiment we 
gradually cross from superconducting to non-superconducting regions
according to the phase diagram of fig.1. Some such typical traces are
shown in the figure --- the dashed lines OA, OXB and OC. Traces OXB 
and OC represent a superconductor - insulator transition, a case
which has been discussed very widely in the literature\cite{seb}. On the
other hand, the trace OA represents a superconductor - metal - insulator
transition.  In this part of the phase diagram, the retardation effects 
are very strong and the system passes through an intermediate 
disordered phase. This is a new prediction made by our analysis. Model (1) 
which leads to this is good for s-i-s junctions.

\section{Resistivity of the metallic phase}

The discussion in the previous section suggests that the appropriate model for
the Bose metal phase is that of uncondensed bosons (vortices and 
antivortices) in a transverse gauge field.
In the frustrated BM phase, the vortices are unbound, but they fail to
bose condense due to retardation, or equivalently dissipative, effects. As a 
result, the system 
exhibits metallic behaviour. This section is devoted to making 
a simplest possible estimate of the resistivity of this metallic phase.
We shall evaluate the resistivity of the charge bosons
$\rho_{c}$ in terms of the conductivity $\sigma_{v}$ of the dual variables,
viz. the vortices. The relation between the two is given by\cite{mpf}
(we mention the factor of $h$ explicitly in this formula) ---
\begin{equation}
\rho_{c} = (h/4e^{2})\sigma_{v}
\end{equation}
Also, $\sigma_{v}$ is given by the Drude formula $\sigma_{v}=n_{vf}\tau_{
tr}/m_{v}$, where $\tau_{tr}$ refers to the transport time and $n_{vf}$
is the free vortex density. There are 
three contributions to the conductivity of the vortices ---
(i) dissipation due to the action of a heat bath(coming from the trasverse
modes),
(ii) scattering from impurities, and
(iii) Bardeen-Stephen processes.
In this part, we shall focus on the contribution from process (i).  We 
give here a short description of how this dissipation mechanism comes 
about in this BM phase. As one enters this phase from the SC phase, say
(following curve A in fig.1, for example), the vortices and antivortices
unbind in 2-d. These unbound charges screen each other leading to a
finite screening length, i.e. a gap in the longitudinal part of the
gauge field. 
However, the transverse part of the gauge field
is gapless, because there is no spontaneous symmetry breaking. These
transverse modes have their source in quantum fluctuations and mediate
the dissipation process. They represent the plasmons. (Physically, this 
makes sense, since the plasmons are gapless in 2-d.) The physics is 
essentially
that of damping which results when a charge (vortex) moves through a 
background of other charged particles. This lends a hydrodynamical
character to this dissipation mechanism.
In section IId, we saw that the most important modes are $\omega\ll c_{s}k$.
So, we shall consider contributions from these modes only. The leading order 
contribution, then, comes from the $\omega = 0, k=$ finite modes, and we
make an estimate from this sector only. We think the higher order
contributions will lead to simple renormalisation of coefficients, so 
long as the density of vortices (and antivortices) is low (please see
below). Thus, we are led
to evaluate the resistivity of a set of uncondensed bosons moving in
a static random (but, annealed) gauge field, with a variance
\[ <H(q)H(-q)> = 1/2c_{s}^{2}ln(2/J\Delta\tau)   \]
as seen from eqn.(15),
where $H = \vec{\nabla}\times\vec{A}$. Now, also in this delocalised phase,
the interactions between the vortices are screened (by antivortices)
and, in a lowest order approximation, we shall treat them as 
non-interacting bosons. Thus,
the transport time is the same as that of a non-interacting particle
scattering from a random magnetic field, which is given by\cite{nal} ---
\[ \tau_{tr} = \frac{m_{v}}{\pi^{2}}. 2(V_{0} + 8V_{1})\Delta\tau   \]
where we have used the above field distribution and the parameters as
enumerated following eqn.(13).
The lattice constant in the time direction, $\Delta\tau\sim 1/\Delta_{0}$
\cite{bd}. Thus, we have ---
\begin{equation}
\rho_{c}\sim \frac{2}{\pi^{2}}R_{Q} (n_{vf}\xi_{0}^{2})(V_{0} +
8V_{1})/\Delta_{0}
\end{equation}
where $R_{Q}=h/4e^{2}=6.45K\Omega$.
The quantity $n_{vf}\xi_{0}^{2}$ is
inversely proportional to the square of the superconducting coherence length 
which diverges at the SC-BM phase boundary, i.e. $n_{vf}\xi_{0}^{2} \sim
(\xi_{0}/\xi_{+})^{2} \sim [(V_{0}+8V_{1})/\tilde{b}_{0}J - 
1]^{2\nu}$, obtained from eqn.(17). Using eqn.(3), we can rewrite this as ---
\begin{equation}
 n_{vf}\xi_{0}^{2} \sim (\frac{R_{n}}{R_{c}}-1)^{2\nu}, 
\end{equation}
where $R_{c}=\tilde{b}_{0}R_{Q}/
[(V_{0}+8V_{1})/\Delta_{0}]$ is the critical resistance of the film
where the metallic phase sets in and $\nu=$ correlation length exponent. The
exponent $\nu$ is dependent on the particular universality class of
SC-BM phase transition. For our 
case, the SC-BM phase boundary has the character of a (2+1)-d XY model
and, hence, $\nu \simeq 2/3$. However, a real material, like the low-$T_{c}$
thin film systems, is heavily disordered and $\nu$ will certainly be very
different from this, as we shall see below. So, from eqn.(21), we have 
\begin{equation}
R_{\Box}/R_{Q} \sim \tilde{b}_{0}[(R_{n}/R_{c})-1]^{2\nu}(R_{Q}/R_{c})
\end{equation}
{\em The case when disorder is present---} 
The foregoing discussion can be readily generalised to the case when disorder
is present. Owing to the Drude formula, the structure of eqn.(21) is not 
unique to model (1) but will come about when the bosons form a metallic phase
as a result of phase fluctuations (in zero magnetic field), i.e.
\begin{equation}
R_{\Box} \sim 2R_{Q} (n_{vf}\xi_{0}^{2}) \mu_{D}
\end{equation}
where $\mu_{D}$ represents the vortex mobility, or the friction factor,
arising from the dissipation of vortices and is model dependent\cite{n7}.
For the pure case discussed above, the damping is due to the transverse
modes
and $\mu_{D}\approx (V_{0}+8V_{1})/\pi^{2}\Delta_{0}$. When the normal
resistance is tuned through $R_{c}$, as we saw above for the pure case,
the most important dependence of BM resistivity in the formula (24) comes
from the free vortex density $n_{vf}\xi_{0}^{2}$, i.e.
\[ n_{vf}\xi_{0}^{2} \sim (\frac{R_{n}}{R_{c}}-1)^{2\nu}  \]
where the critical exponent $\nu$ characterising the SC-BM phase 
transition depends on the particular universality class being considered. As
compared to this, the dissipation factor $\mu_{D}$, in general being 
dependent on the morphological and normal properties of the film, is only
weakly dependent on the normal resistivity about $R_{c}$\cite{n7}. Thus,
in general,
\begin{equation}
R_{\Box}\sim R_{Q}[(R_{n}/R_{c})-1]^{2\nu}.
\end{equation}
For the case of Ref.\cite{ag}, $R_{c} \sim R_{Q}$. As $R_{n}$ is tuned in such
an experiment, one executes a typical trajectory A on the phase diagram 
of fig.1. The phase which exists in a real material like Gallium and is
not captured by a pure model like model (1) is the Bose glass(BG) phase. 
So, instead of SC-BM-BI scenario along A, one would probably have a 
SC-BM-BG(-BI) scenario. Either way, in the BM phase, because of formula (25)
the resistivity of this metal increases continuously 
from zero through a wide spectrum of values as $R_{n}$ is tuned through 
$R_{c}$, as is seen in the experiments\cite{ag,agb}. We display the 
low temperature metallic resistivity from Ref.\cite{ag}[fig.2
of this reference] as a function of
$R_{n}$ in fig.2 and observe the remarkable fit to the formula (25) with
$\nu\simeq2$. This clearly points out that the low temperature metallic 
phase observed in Ref.\cite{ag} was, in all probability,
due to an incoherent motion of the Cooper pairs.
The aforementioned value of the correlation exponent is consistent with the
notion that Ga films were highly disordered samples\cite{tvr} and the Chayes'
theorem\cite{cha} along with it.

\section{How new is this state of matter?}

The question now arises whether the Bose metal discussed in the previous 
sections is actually an adiabatic continuation of some known state of
matter or if it is completely new. In this section, we argue that this state
is actually a mathematical variant of a quantum disordered spin liquid
with {\em large} spin and that it is a liquid of non fermionic variety.

To show this, we first note that the algebra which controls the quantum
mechanics of our problem is $E_{2}$, as discussed in section II.
Now, $E_{2}$ is a group contraction of $SO(3)$ \cite{e2}, i.e. as the 
radius $R$ of the sphere on which $SO(3)$ operations are defined tends
to infinity, $SO(3)$ contracts to $E_{2}$, viz. $J_{x}/R\rightarrow -P_{y},
J_{y}/R\rightarrow P_{x}, J_{z}\rightarrow L$, where the operators
$L$ and $P$ are as defined in eqn.(5). And, in the same limit, i.e.
as the SO(3) angular momentum quantum number $j = pR \rightarrow \infty$,
where $p$ refers to the linear momentum quantum number of $E_{2}$, 
the irreducible representations of $SO(3)$ map onto those of $E_{2}$.
Thus, the model (1) is a variation of a quantum spin model and the
Bose metal obtained here, which is a quantum disordered phase, may be
thought of as analogous to the frustrated spin liquids obtained within
such a model in the limit of spin becoming large\cite{douc}.

A more intriguing question is the issue of how consistent the existence of
a disordered phase as that of a Bose metal at $T=0$ is with the third law
of thermodynamics. To this end,
we calculate the low temperature specific heat of the metal under the same
approximation as in section IV, i.e. non-interacting uncondensed bosons
(vortices) moving in a transverse gauge field. The longitudinal part
of the gauge field is gapped because of the screening effects and hence 
do not contribute significantly to the specific heat\cite{tsv}. 
However, the transverse gauge field representing the quantum 
fluctuations associated with the transverse modes are gapless
in this metallic phase\cite{il} (because there is no spontaneous symmetry
breaking) and, hence, make a dominant contribution to the specific
heat. The calculation for uncondensed bosons interacting with a transverse 
gauge field\cite{kalm,nal} is analogous to the fermionic case
for which an extensive literature exists\cite{tsv}. Following 
Reizer\cite{tsv}, we find at low temperatures (Appendix E)
\begin{eqnarray}
    C &\sim& A_{p}T^{2/3} \textup{\hspace{0.5cm} (pure)} \nonumber \\
      &\sim& A_{d}Tln(T_{0}/T) \textup{\hspace{0.5cm} (with disorder)}
\end{eqnarray}
This anomalous behaviour originates sheerly from the fact that the 
transverse gauge field undergoes dynamical screening, which leads to the
presence of diffusive modes $\omega \sim -iq^{n}$, where $n = 2$(disordered)
or $3$(pure). The coefficient $A_{p}$ scales as $A_{p}\sim n_{f}^{2/3}$,
$A_{d}$ as $A_{d}\sim n_{f}$, and $T_{0}$ as $T_{0}\sim 1/n_{f}$, where
$n_{f}$ refers to the free vortex density (Appendix E). This displays the
non-fermi liquid behaviour of the Bose metal phase and verifies that
there is no consistency problem with the third law. Clearly, the BM phase has
more entropy than a normal fermi liquid at low temperatures.

The foregoing temperature variation of the specific heat shows that there
are weak singularities at $T=0$ in the metallic state. Thus, the BM state
has the character of a {\em critical} point as $T\rightarrow 0$. This requires
further study since 
the coefficients which enter the specific heat calculation would undergo 
strong renormalisation and the actual temperature dependence may be somewhat 
different. These arguments also suggest that there
is no good length scale in the quantum liquid phase at $T=0$. This can be
seen by calculating
the charge charge correlation function. Using equations (A4) (or, equivalently
(B2)) and (E3), we obtained:
\begin{equation}
<\delta n_{q,\omega}\delta n_{-q,-\omega}> \approx \frac{q^{2}}{-i\tilde{a}
\frac{\omega}{q} + \tilde{b}q^{2}}
\end{equation}
where the coefficients $\tilde{a}$ and $\tilde{b}$ may be read off from
eqn. (E3). Thus, the excitation spectrum of the charge fluctuations is
gapless and diffusive. As we pointed out in section IIB, because of the
algebra of phase fluctuations, this does not 
contradict the fact that there is a length scale associated with 
superfluidity viz. the
superfluid correlation length $\xi_{+}$ tied to the free vortex density
$n_{f}$. Thus, the BM phase is analogous to
a gapless quantum spin liquid, very much like an RVB state.

\section{Discussion and conclusions}

In summary, we have argued that the superconductor-insulator transition in
low$-T_{c}$ superconductors should be viewed as a {\em two} order parameter
problem and observed that it is described by two coupled
XY models in (2+1)-d in the charge picture. This change
from a single (2+1)-d XY model, as conventionally thought, is tied to
the inclusion of an appropriate vortex mass term in the vortex picture or 
non-local interactions in the charge picture. This
leads to the interesting possibility of a novel disordered Bose metal phase
distinct from the traditional superconductor and Bose insulator phases. 
On the basis of our analysis of the model, we expect a superconductor -
insulator transition when $V_{on-site}\gg V_{n.n.}$ ($n.n.=$ nearest 
neighbour), a case close to that of the Josephson Junction arrays (with
$B_{ext} = 0$); and a superconductor - metal - insulator transition
when $V_{on-site}\sim V_{n.n.}$, a situation close to that of the 
granular superconductors (it may be helpful to note that $V_{i}\sim
(2e)^{2}/C_{i}$, $C_{i}=$ capacitance)\cite{jem}. The physics of the
problem seems to be controlled by a multicritical point. We
find the properties of the Bose metal
to be critical. It is inherently connected to the quantum
spin liquid states and is strongly reminiscient of the gapless RVB phase.
The resistivity of this novel metallic phase predicted by our calculations
finds an excellent match with the experiments on Gallium films\cite{ag}.
Dissipation in the bosonic system at low temperatures, within our model,
is hydrodynamical: it
comes from the fact that a moving charge (vortex) dissipates
as it moves through a background of charged particles (vortices and 
antivortices). The heat is carried away by the  gapless transverse modes 
representing the plasmons.
This source of dissipation translates into quantum fluctuations
in the charge picture.

To our knowledge, so far there has been no metallic phase proposed in the phase 
diagram of the Bose localisation problem, within the scope of any physically
realisable
model. This paper is a first attempt to propose this concept and establish
some of the basic principles which underlie the existence of such a phase.
As a result, many of the features which are tied to the current experiments
have gone unaddressed. Some of these are: existence of disorder, 
long-range $1/r$ interactions, non-zero magnetic field, finite temperature
effects, etc. Also, in a realistic sample, an important source of heat bath
in addition to the emission of transverse excitations is
the electrons in the vortex core. This needs to be incorporated into the
calculation.

An important concern is whether the BM phase will survive in improved 
calculations, given that our calculations have been mean field like. As
we saw, the experiments argue in
favour of the existence of this phase. Further, as we argued in section IId,
the retardation effects leading to the destruction of vortex superfluidity
and the binding of vortex - antivortex pairs, the features corresponding
to the loss of charge order and the growth of phase order respectively,
are controlled effectively by two separate parameters. Plus,
as we found in section III, the constants which determine the curves LXM
and ZYX in fig.1 differ by about an order of magnitude because of the strong
retardation effects felt by the vortices in this model. In view of these
facts and the self consistent nature of the calculation, it is reasonable
to think that the curve ZYX will lie above the curve LXM and intersect it
at a multicritical point even if the fluctuations are taken into account,
implying a separate BM phase\cite{n6}. Either way,
it will be very useful to check this via numerical simulations.
An important issue which confronts us at the moment is ---
what is the universality class of the critical phenomena 
associated with model (1)?
Although we have commented on this above, one needs to perform renormalisation
group calculations or numerical simulations to 
answer this question completely. What is clear from our
calculations, however, is that the critical behaviour, both at $T=0$ and
$T=$ small but finite, implies a new universality class with a 
multicritical point which is very rich.

Given the considerable amount of work done on the Bose-Hubbard models, one
might wonder why this phase was not observed in the other theoretical
constructs, particularly in simulations. In view of this, we would like to
mention a few points, in addition to the idea of two order parameters, 
which distinguishes our work from the previous ones. Firstly,we have worked 
in the large $n_{0}-$limit of the problem. Most of the current work on this
model does not treat this limit consistently. Also, in this limit, the 
algebra which
determines the quantum mechanics is qualitatively different from that of the
hard core bosons, i.e. $E_{2}$, and 
the chemical potential $\mu$ does not play a significant role (as found
in section IIb). In the small
$n_{0}-$limit, $\mu$ plays a dominant role, especially in stabilising the
commensurate Bose insulator(CBI) phase (mostly called Mott insulator in the 
Bose literature). The weakening effect of $\mu$ as $n_{0}$ increases
is clearly observable in the shrinking of CBI lobes with $n_{0}$,
evidenced by the perturbative and Quantum Monte Carlo calculations of
Ref.\cite{mon}. This provides additional support to the results obtained here in
this approximation. 
Further, we also observed that the nearest neighbour
interaction plays a crucial role in opening up the metallic phase in the
system. There is no such phase with just on-site repulsion. This piece of 
physics
has support from the RG flows constructed by Fisher and Grinstein\cite{fg}.

{\em On the supersolid phase}: It is somewhat tempting to use eqn.(19) 
to the fullest extent and
extend the curve ZYX beyond the point X along the dotted line XO. Eqn.(19)
is really not good beyond the point X; but if we take it at face value,
it seems to suggest a region OXM where both phase order and charge order
exist, i.e. a supersolid phase. This has some support, if we look at the
action (10). Here, as we noted earlier, when $V_{1} \ll V_{0}$,
$\theta_{i} \simeq 0$, implying the presence of charge order in the
system. This, however, corresponds to trivial charge ordering only,
viz. $\delta n_{i}\simeq 0$, i.e. $n_{i}\simeq n_{0}$. This phase needs 
to be verified by better calculations. If this 
phase exists in improved calculations, the point X would be tetracritical.

Before we close, we would like to make a short comment on the Mo-Ge 
system\cite{kgb}, a situation where magnetic field is present. As we argued
in section IId, the physics of quantum phase fluctuations in 2-d 
superconductors is a 
quantum mechanical generalisation of the KT scenario. We expect this to 
happen when magnetic field is present as well. Let us recollect the 
classical case for completeness. Here, dislocation - antidislocation pairs
are induced in the vortex lattice which unbind as a result of 
thermal fluctuations \cite{dhu}. At low temperatures, quantum effects dominate
and we expect the unbinding resulting from quantum fluctuations instead.
But, as pointed out in this paper, their kinetic energy should
receive two contributions --- from the zero point motion and the action of a
heat bath. The metallic state is realised when the dislocations in the 
vortex lattice(more precisely, vortex glass) phase move under the 
action of a heat bath. This is very much along the lines of what had
been suggested in Ref.\cite{kgb}. Following our discussion in section 
V, we suggest 
specific heat measurements be made to spot any non-fermionic behaviour 
of this metal-like phase. We plan to take up some of the unresolved issues
mentioned in this section in a future publication.

\acknowledgments

This work was supported by NSF under grant no. DMR 96-27459.

\appendix
\section{The duality transformation}

We start from action (9), but with the transformation $m_{i}\rightarrow
e^{-iQ.r_{i}}m_{i}$ undone, so that 
\begin{eqnarray}
S&=&i\sum_{i}m_{i}(\nabla_{\tau}\phi_{i}) + V_{0}\Delta\tau
\sum_{i}m_{i}^{2} \nonumber \\&+&V_{1}\Delta\tau\sum_{i\alpha}(m_{i} +
m_{i+\alpha})^{2} - J\Delta\tau\sum_{i\alpha}cos(\nabla_{\alpha}\phi_{i}).
\end{eqnarray}
We now use Villain transformation on the $J$ term. For small values of
$J\Delta\tau$, one has\cite{bd}
\[ e^{J\Delta\tau\sum_{i\alpha}cos(\nabla_{\alpha}\phi_{i})} \simeq
e^{-ln(\frac{2}{J\Delta\tau})\sum_{i\alpha}n_{i\alpha}^{2} +
i\sum_{i}\phi_{i}(\overline{\nabla}_{\alpha}n_{i\alpha})}  \]
Integrating out the phase degrees of freedom $\phi_{i}$, one obtains ---
\[ Z = \sum_{\{m_{i}\}}\sum_{\{n_{i\alpha}\}}e^{-S}, \]
\begin{eqnarray}
S&=&ln(\frac{2}{J\Delta\tau})\sum_{i\alpha}n_{i\alpha}^{2} +
V_{0}\Delta\tau\sum_{i}m_{i}^{2} \nonumber \\ & + & 
V_{1}\Delta\tau\sum_{i\alpha}(m_{i} +m_{i+\alpha})^{2}
\end{eqnarray}
supplemented with the constraint 
\begin{equation}
\overline{\nabla}_{\tau}m_{i} + \overline{\nabla}_{\alpha}n_{i\alpha} = 0,
\end{equation}
(with the summation over $\alpha$ implied). Defining $n_{i}^{\mu}=(m_{i},
n_{i}^{\alpha})$, $(\mu=0,1,2)$, one obtains the constraint (A3) as ---
\[ \overline{\nabla}_{\mu}n_{i}^{\mu} = 0. \]
This allows us to define an integer gauge field $A_{i}^{\mu}$, via
\begin{equation}
n_{i}^{\mu} = \varepsilon_{\mu\nu\rho}\overline{\nabla}_{\nu}A_{i}^{\rho}.
\end{equation}
A purist might object to our not using the shifted lattice operators
$A_{i-\hat{\rho}}^{\rho}$ in eqn.(A4)\cite{it}. This does not matter for 
our case since the model involves a single gauge field. Now,
transforming over to the Fourier space ($m_{i}=\frac{1}{\sqrt{N_{0}}}
\sum_{q,\omega}e^{i\tilde{q}.x_{i}}m_{q,\omega}$, with $\tilde{q}=
(\omega,\vec{q})$ and $x_{i}$ referring to the $i^{th}$ site on the 
space-time lattice, with $N_{0}$ the total no. of lattice sites), one
obtains,
\[ Z = \sum_{\{A_{i}^{\mu}\}}e^{-S}, \]
with
\begin{equation}
S=ln(\frac{2}{J\Delta\tau})\sum_{q,\omega}[\hat{q}^{2}A^{0}_{q,
\omega}A^{0}_{-q,-\omega} + G^{-1}(\omega,q)A^{\alpha}_{q,\omega}
A^{\alpha}_{-q,-\omega}]
\end{equation}
with $G^{-1}(\omega,q)$ given by eqn.(12) and $A_{q,\omega}^{\mu}$ refer to
the fourier transform of $A_{i}^{\mu}$. Here $\hat{q}$ and $\hat{\omega}$ refer to 
lattice momentum and frequency respectively, and $\hat{q}^{2}=\sum_{\alpha}
\overline{K}_{\alpha}(q)K_{\alpha}(q)$, with 
$\overline{K}_{\alpha}(q)=\frac{1}{i}
(1-e^{-iq_{\alpha}}), K_{\alpha}(q)=\frac{1}{i}(e^{iq_{\alpha}}-1)$ \cite{ld}.
And, $(\hat{q}-Q)^{2} = \sum_{\alpha}\overline{K}_{\alpha}(q-Q)
K_{\alpha}(q-Q)$ in eqn.(12).
Eqn.(14) is complete with the gauge condition,
\begin{equation}
{\nabla}_{\alpha}A_{i}^{\alpha} = 0.
\end{equation}
Now, we introduce the vortex variables $j^{\mu}_{i}$ via the Poisson 
summation formula\cite{it} ---
\begin{equation}
Z = \sum_{\{j_{i}^{\mu}\}} \int{\mathcal D}A_{i}^{\mu}e^{-S -
2\pi i\sum_{i}j_{i}^{\mu}A_{i}^{\mu}}.
\end{equation}
Integrating out the gauge field degree of freedom, one obtains eqn.(11).

\section{Continuum Version}

The features of the vortices discussed in section IId are not special to
the lattice case, but exists in the continuum case as well. This gives us
a more general frame in which to study the model. The
considerations in the continuum case start from the action\cite{scz,dd} ---
\begin{eqnarray}
\tilde{S} =\int d\tau d^{2}x[(m/2\bar{\rho})\mid\vec{J}\mid^{2}&+&\frac{1}{2}
\int d^{2}y V(x-y)\delta\rho(x)\delta\rho(y) \nonumber \\
& + & 2\pi i \tilde{j_{\mu}}A_{\mu}]
\end{eqnarray}
where $\vec{J}$ refers to the charge current
and $\delta\rho$ the charge fluctuations, with the gauge
field $A_{\mu}$ given by ---
\begin{equation}
J^{\mu}(x)=(\delta\rho,\vec{J})=\varepsilon_{\mu\nu\rho}\partial_{\nu}A_{\rho}
\end{equation}
Also, $\tilde{j_{\mu}} = (\rho_{v},\vec{J_{v}})$ refers to the vortex three
current density, $m=$ mass of the Cooper pairs, and $\bar{\rho}=$ average
Cooper pair density. We have not written down a Magnus force term on 
vortices due to a background charge condensate, because this is not
contained in the lattice model (1) discussed in the paper. However, this
may be important in a uniform system like Mo-Ge\cite{akr}.
Working in the transverse gauge $\vec{\nabla}.\vec{A}=0$, we obtain
after integrating out the gauge field ---
\begin{eqnarray}
\tilde{S}&=&2\pi^{2}\frac{\bar{\rho}}{m}\sum_{k,\omega}
\tilde{j}^{0}_{k,\omega}\frac{1}{k^{2}}\tilde{j}^{0}_{-k,-\omega} \nonumber \\
& + & 2\pi^{2}\frac{\bar{\rho}}{m}\sum_{k,\omega}
\tilde{j}^{\alpha}_{k,\omega}\tilde{G}(\omega,k)\tilde{j}^{\alpha}_{-k,-\omega}
\end{eqnarray}
where the Green's function $\tilde{G}_(\omega,k)$ is given by ---
\begin{equation}
\tilde{G}^{-1}(\omega,k) = \omega^{2} + (\bar{\rho}/m)k^{2}V_{k}
\end{equation}
where $V_{k}$ is the Fourier transform of the potential $V(r)$. From here
onwards we consider the screened Coulomb potential with
\[ V_{k} = V_{0}\alpha/\sqrt{k^{2}+\alpha^{2}}  \]
In the limit $\omega,k\rightarrow 0$ and $\omega\ll\tilde{c_{s}}k$ (with
$\tilde{c_{s}}$ being the plasmon velocity), the Green's function splits
up into a singular part $\tilde{G_{s}}(\omega,k)$ and $\tilde{G_{0}}$ just
like in the lattice model ---
\begin{equation}
\tilde{G}(\omega,k) \simeq \tilde{G}_{s}(\omega,k) + \tilde{G}_{0},
\end{equation}
where
\[ \tilde{G}_{s}(\omega,k)=1/(\omega^{2}+\tilde{c}_{s}^{2}k^{2}) \textup{, }
\tilde{G}_{0} = 1/(2\bar{\rho}V_{0}\alpha^{2}/m),  \]
with
\[ \tilde{c}_{s}^{2} = \bar{\rho}V_{0}/m.  \]
Thus, from (B3) and (B5), we obtain ---
\begin{eqnarray}
\tilde{S}&=&2\pi^{2}\frac{\bar{\rho}}{m}\sum_{k,\omega}
\tilde{j}^{0}_{k,\omega}\frac{1}{k^{2}}\tilde{j}^{0}_{-k,-\omega} \nonumber\\
& + & 2\pi^{2}\frac{\bar{\rho}}{m}\sum_{k,\omega}
\tilde{j}^{\alpha}_{k,\omega}(\tilde{G}_{s}(\omega,k) + \tilde{G}_{0})
\tilde{j}^{\alpha}_{-k,-\omega}
\end{eqnarray}
Thus, comparing eqns.(B6) and (14), we see that the lattice and continuum
actions of phase fluctuations, within the scope of our model, have identical
low frequency and long wavelength limits.

Correspondence between the lattice version and the continuum version is
established via the following identifications ---
\begin{eqnarray}
m/\bar{\rho}&\longleftrightarrow&2ln(2/J\Delta\tau) \\
\tilde{c}_{s}^{2}=\bar{\rho}V_{0}/m&\longleftrightarrow&c_{s}^{2} = 
(V_{0}+8V_{1})\Delta\tau/ln(2/J\Delta\tau) \\
\tilde{G}_{0} = 1/(2\bar{\rho}V_{0}\alpha^{2}/m)&\longleftrightarrow&
G_{0} = b^{2}/8(b^{2}+c^{2})^{2}
\end{eqnarray}
Following Ref.\cite{scz}, one can easily show that eqn.(B3) is equivalent to the
following Hamiltonian for the vortices ---
\begin{equation}
\tilde{H}_{v} =\frac{1}{2m_{v}}\sum_{i}(\vec{p}_{i}+2\pi q_{i}\vec{A}(x_{i}))^{2}
+ 2\pi^{2}\frac{\bar{\rho}}{m}\sum_{i\neq j}q_{i}q_{j}ln\mid r_{i} -
r_{j} \mid
\end{equation}
where the summation is over all vortices and antivortices and the gauge 
field $\vec{A}$ has the spectrum determined by $\tilde{G}_{s}^{-1}(\omega,k)$
as in eqn.(15). Here $q_{i}=\pm 1$ refers to the charges on the vortices.
Here $m_{v}$ refers to the vortex mass.

\section{An estimate of $G(0)$}

From section IIIa, we have ---
\begin{equation}
 G(0) = \frac{1}{2}\int^{\pi}_{-\pi}\frac{d\omega}{2\pi}
	\int^{\pi}_{-\pi}\frac{d^{2}q}{(2\pi)^{2}}
	\frac{1}{(1 - cos\omega) + A^{2}}
\end{equation}
where 
\begin{equation}
A^{2} = [2 - (cos q_{x} + cos q_{y})][\kappa_{0}^{2} + 2\kappa_{1}^{2}
	\{2 + cos q_{x} + cos q_{y} \} ]
\end{equation}
with
\[ \kappa_{0}^{2} = c^{2}  \textup{ and } \kappa_{1}^{2} = b^{2}/8, \]
where $b^{2}$ and $c^{2}$ have been introduced in section IId.

From (C1), we obtain 
\begin{equation}
 G(0) = \frac{1}{2}\int^{\pi}_{-\pi}\frac{d^{2}q}{(2\pi)^{2}} F(q),
\end{equation}
with
\begin{equation}
 F(q) = \frac{1}{\sqrt{A^{2}(2 + A^{2})}}
\end{equation}
Important contributions to (C3) come from the low momentum region. Thus,
\begin{equation}
 F(q) \simeq 1/\sqrt{2(\kappa_{0}^{2} + 8\kappa_{1}^{2})}
	 \sqrt{2 - (cos q_{x} + cos q_{y})}
\end{equation}
Combining eqns.(C3) and (C5), we are led to the condition (17), with
\begin{equation}
 b_{0} = \int_{0}^{\pi/2}\frac{d^{2}x}{\sqrt{sin^{2}x+sin^{2}y}}
\end{equation}

\section{Self consistent functional approach}

In this appendix, we give a brief review of the self-consistent functional 
approach\cite{pop} discussed by Ioffe et. al.\cite{il}. This is essentially 
a hydrodynamical kind of approach. We start with the following
electromagnetic Lagrangian of bosons in imaginary time (in $2-d$) ---
\begin{equation}
L_{B} = L_{B}^{0} + \tilde{L}_{B} + L_{B}^{a}.
\end{equation}
Here
\begin{eqnarray*}
 L_{B}^{0} = \psi^{\ast}\frac{\partial\psi}{\partial\tau}
	+ \frac{1}{2m}\mid\vec{\nabla}\psi\mid^{2} 
&+& \frac{g^{2}}{2}\rho(r)ln\mid r - r'\mid\rho(r') \nonumber \\
&-&\mu\mid\psi\mid^{2}    \textup{\hspace{0.5cm}} -(D2a) 
\end{eqnarray*}
\[ \tilde{L_{B}} = -\frac{i}{2m}\vec{a}.[\psi\vec{\nabla}\psi^{\ast}
	- \psi^{\ast}\vec{\nabla}\psi] + \frac{1}{2m}\vec{a}^{2}
	\mid\psi\mid^{2} \textup{\hspace{0.5cm}} -(D2b) \]
\[ L_{B}^{a} = \frac{1}{2g^{2}}[(\partial_{\tau}\vec{a})^{2}
		+ c^{2}(\vec{\nabla}\times\vec{a})^{2}] 
	\textup{\hspace{0.5cm}} -(D2c) \]
where $\psi, \psi^{\ast}$ refer to the Bose fields and $\rho=\psi^{\ast}
\psi$. We work in the transverse gauge $\vec{\nabla}.\vec{a}=0$.
The dimensionless coupling constants determining the strength of the
Coulomb repulsion and transverse gauge field, viz. $\alpha_{c}$ and
$\alpha_{g}$ respectively, are defined as ---
\setcounter{equation}{2}
\begin{eqnarray}
\alpha_{c} &=& \frac{g^{2}m}{16\pi^{2}n} \nonumber \\
\alpha_{g} &=& \frac{g^{2}}{8\pi mc^{2}},
\end{eqnarray}
where $n$ is the average density of bosons. Comparison of eqn.(D2) with
eqn.(15) allows us to identify ---
\begin{eqnarray}
 g^{2} &\leftrightarrow& 2\pi^{2}/ln(2/J\Delta\tau) \nonumber \\
 c^{2} &\leftrightarrow& c_{s}^{2} \nonumber \\
\textup{and, \hspace{0.5cm}} m/2n &\leftrightarrow&
	\pi^{2}G_{0}/ln(2/J\Delta\tau)
\end{eqnarray}
The last identification follows from the fact that the weight of the
zero point motion term $\mid\vec{j}\mid^{2}$ is $m/2n$. Using eqns.
(D3) and (D4), we are led to eqns.(18)\cite{n5}.

The calculation of the depletion of the Bose condensate from the action
(D2) proceeds in two steps. The first involves determining the effective
functional of the superfluid Bose system in the absence of a transverse
gauge field. And, the second involves determining the effect of the gauge
field on this system, i.e. calculating the change in the superfluid
density. 

The effective functional of the superfluid Bose system (with $\vec{a} = 0$)
is obtained by expanding the Bose field in terms of the slow and fast 
modes, and integrating out the fast modes\cite{pop}. Galilean invariance
of this system dictates $n_{s} = n$\cite{pns}. The leading terms
in the effective action are (for small values of $\alpha$)---
\begin{eqnarray}
S_{0} = \int d^{2}x d\tau [\frac{n}{2m}\mid\vec{\nabla}\phi\mid^{2}
&-& i\pi\partial_{\tau}\phi + \frac{1}{8mn}\mid\vec{\nabla}\pi\mid^{2}]
\nonumber \\ &+& \frac{g^{2}}{2}\sum_{q,\omega}\pi_{q,\omega}
	\frac{1}{q^{2}}\pi_{-q,-\omega}
\end{eqnarray}
where $\phi$ and $\rho(x,\tau)=n + \pi(x,\tau)$ are the phase and the
amplitude of the slow mode $\psi_{0}=\sqrt{\rho(x,\tau)}e^{i\phi(x,\tau)}$.
We now work in the real time and integrate out the phase variables.
This leads to an action solely in terms of the density fluctuation
variables $\pi$ ---
\[ e^{\frac{i}{2}\int d^{3}xd^{3}x'\pi(x)C^{-1}(x-x')\pi(x')} \]
where 
\begin{equation}
C(\omega,q) = \frac{nq^{2}/m}{\omega^{2} - (q^{2}/2m)^{2} - ng^{2}/m 
	+ i\delta}
\end{equation}
and $d^{3}x = d^{2}x dt$. 

To calculate the change of the superfluid density $n_{s}=n$ due to the 
action of the transverse gauge field, we focus on the diamagnetic 
term in (D2b)
\begin{equation}
 -\frac{1}{2m}\vec{a}^{2} \psi_{0}^{\ast}\psi_{0} = -\frac{1}{2m}
\vec{a}^{2}(n + \pi).
\end{equation}
The weight of the $\vec{a}^{2}$ term determines the superfluid density. One
now integrates out the $\pi$-degrees of freedom. The coupling term in
(D7) leads to an $\vec{a}^{2}-\vec{a}^{2}$ interaction term. One then 
(a) decouples
this term in a mean field approximation and (b) rewrites the gauge
field in terms of fast and slow modes , $a_{1}$ and $a_{0}$, and 
integrates out the fast modes. The weight of the $a_{0}^{2}$ term then 
gives the self-consistent equation for the superfluid density $n_{s}$,
\begin{equation}
n_{s} = n +  \frac{i}{m}\int d^{3}x C(x)D(x)
\end{equation}
where $D(x)$ is the gauge field correlator ---
\begin{equation}
D(\omega,q) = \frac{g^{2}}{\omega^{2} - (c^{2}q^{2} + n_{s}g^{2}/m)+i\delta}
\end{equation}
Eqn.(D8) leads to the phase diagram in $\alpha-\alpha_{g}$ plane, 
mentioned in section IIIb.

\section{Specific heat calculation}

To calculate the specific heat, we consider the Lagrangian (D2) without
the log interation, which is screened in the metallic phase. Connection
with the original model is established via the mapping (D4)\cite{n5}. We first
rescale the transverse gauge field $\vec{a} \rightarrow g\vec{a}$ so that
the effective charge $g$ appears in the eqn.(D2b) rather than (D2c).
Thus, the unperturbed gauge propagator is 
\begin{equation}
D^{0}_{\alpha\beta}(q,\omega_{n}) = K_{\alpha\beta}(q)\frac{1}{\omega_{n}^{2}
	+ c^{2}q^{2}},
\end{equation}
where $\omega_{n}=2\pi nT (n=$integer$)$ are Matsubara frequencies and
\[ K_{\alpha\beta}(q) = \delta_{\alpha\beta} - \frac{q_{\alpha}q_{\beta}}
{q^{2}}. \] 
To obtain the contribution of specific heat from the transverse modes, one
integrates out the particle degrees of freedom in a one-loop approximation
\cite{tsv}. This leads to the retarded polarization function ---
\begin{equation}
\Pi^{(R)}_{\alpha\beta}(q,\Omega) = -K_{\alpha\beta}(q)g^{2}[-i\frac{2n}{m}
\frac{\Omega}{v_{B}q} + c^{2}\chi_{D}q^{2}] 
\end{equation}
where $\chi_{D}=n_{B}(-\mu)/24\pi mc^{2}$ and $mv_{B}^{2}/2 = \mid\mu_{B}
\mid$, with $n_{B}(\xi)=1/(e^{\beta\xi}-1)$. The structure of eqn.(E2) is
quite independent of the statistics because the integrals which enter the
calculation are of the form $\int_{-\mu}^{\infty}d\xi n(\xi)$ and
$\int_{-\mu}^{\infty}d\xi \frac{\partial n}{\partial \xi}$. The 
renormalised gauge propagator is then given by Dyson's eqn. ---
\[ D = D_{0} + D_{0}\Pi D \]
In the quasistatic approximation $\Omega \ll cq$, we have the retarded 
gauge propagator as 
\begin{equation}
 D^{(R)}_{\alpha\beta}(q,\Omega) = K_{\alpha\beta}(q)[1/
	(-i\frac{2ng^{2}}{m}\frac{\Omega}{v_{B}q} + 
	\bar{c}^{2}q^{2})],
\end{equation}
where $\bar{c}^{2} = c^{2}(1 + g^{2}\chi_{D})$. Integrating out the gauge 
fields leads to the free energy ---
\begin{equation}
 \beta F = -\frac{1}{2}Tr ln D(q,\omega_{n})
\end{equation}
The specific heat is obtained from $C = T\partial S/\partial T$. The 
details of the calculation may be found in Ref.\cite{tsv}. For the 
disordered case, the Green's function is written as $G(p,\omega_{n})
= 1/(i\omega_{n}-\xi_{p}+i/2\tau sgn(\omega_{n}))$, with $\tau$ the
scattering time for the vortices from the impurities (which appear as a 
static gauge field in the vortex picture)\cite{dd}. In this case, the
factor $1/v_{B}q$ appearing in the frequency term in (E2) and (E3) gets
replaced by $2\tau$. The coefficients 
$A_{p}$, $A_{d}$ and $T_{0}$ in eqn.(26) are approximately given by ---
\begin{eqnarray}
 A_{p}&\approx&(2ng^{2}/mv_{B}\bar{c}^{2})^{2/3}/\pi \nonumber \\
 A_{d}&\approx& ng^{2}\tau/\pi^{2}m\bar{c}^{2} \nonumber \\
 T_{0}&\approx& \bar{q_{c}}^{2}m\bar{c}^{2}/4ng^{2}\tau
\end{eqnarray}
where $\bar{q_{c}}$ denotes an upper momentum cutoff $\sim 1/\xi_{0}$,
with $\xi_{0}$ the Cooper pair size.

\section{Commutation rule (6c)}

In this appendix, we discuss how the commutation rule (6c) comes about.
From commutators (6a) and (6b), we observe that $P$ and $P^{\dagger}$ are
ladder operators. Hence, in the angular momentum representation
($L\mid\Omega_{m}> = m\mid\Omega_{m}>$)\cite{opt,e2},
\begin{eqnarray}
 P &=&i\sum_{-\{n_{0}\} + 1}^{\infty}\mid\Omega_{m-1}><\Omega_{m}\mid
  \approx i\sum_{-\infty}^{\infty}\mid\Omega_{m-1}><\Omega_{m}\mid \\
P^{\dagger}&=&-i\sum_{-\{n_{0}\}}^{\infty}\mid\Omega_{m+1}><\Omega_{m}\mid
   \approx -i\sum_{-\infty}^{\infty}\mid\Omega_{m+1}><\Omega_{m}\mid 
\end{eqnarray}
where $\{n_{0}\} = n_{0}$ for integer $n_{0}$, and the floor or ceiling
of $n_{0}$ (appropriately taken) otherwise.
The approximate expressions written on the right hold only when $n_{0}\gg 1$.

\newpage
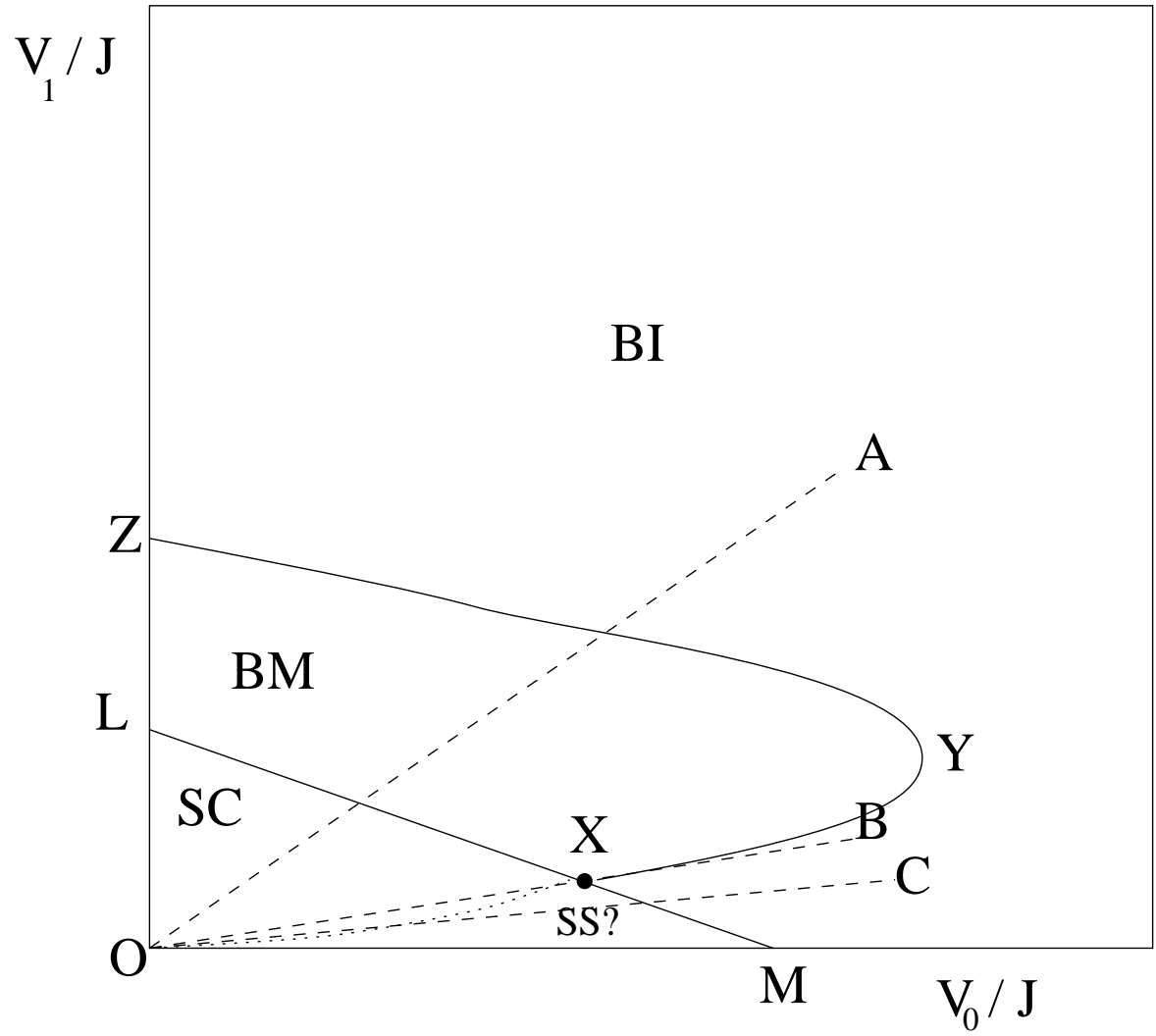
\begin{figure}
\caption{A schematic phase diagram for model (1). The various phases are
demarcated by solid boundaries. X denotes the multicritical point.
The dashed lines (OA, OB, OC) denote typical
trajectories executed as the normal resistance $R_{n}$ of a superconducting
film is tuned. Please see the text for details.}
\label{fig1}
\end{figure}

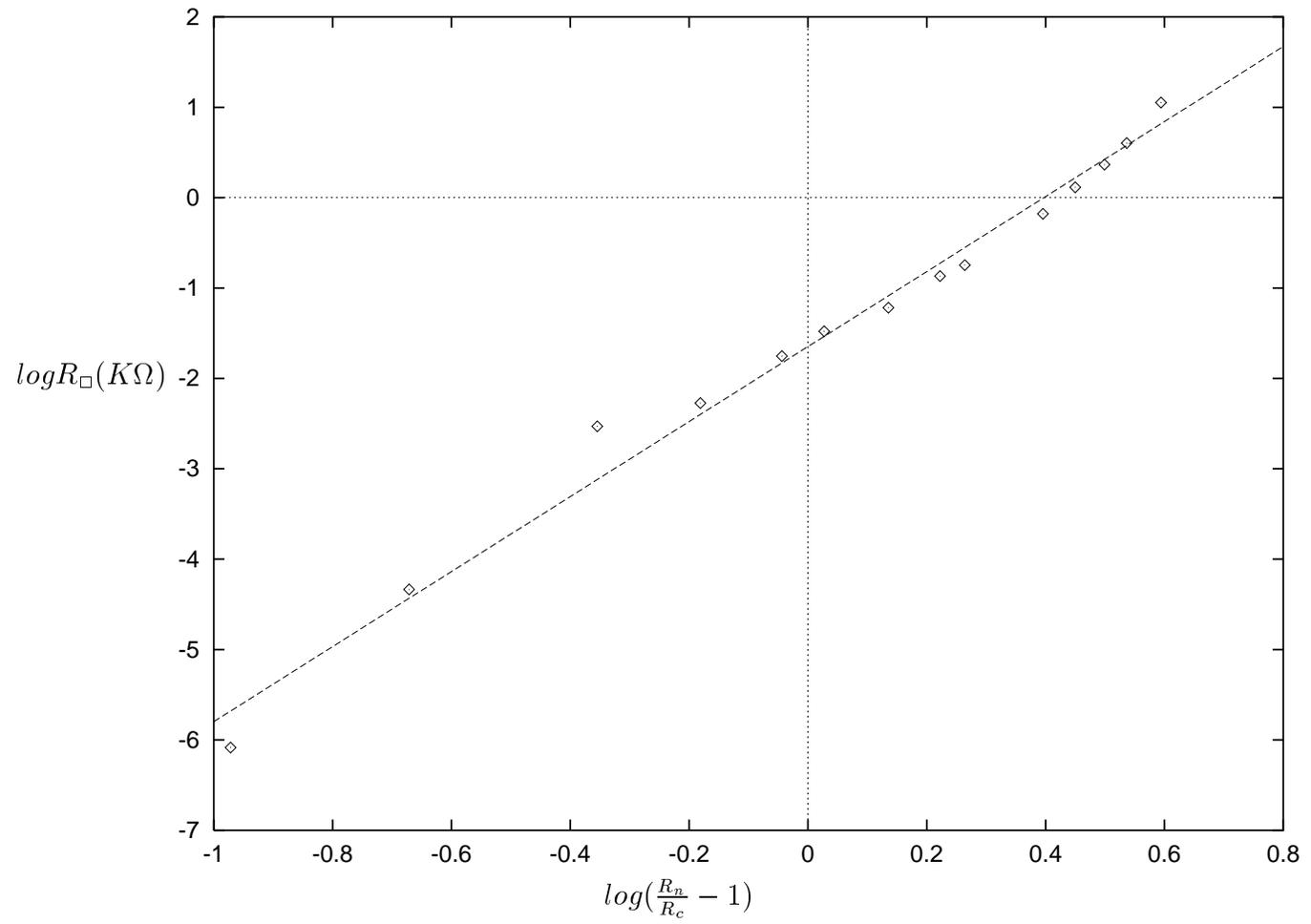
\begin{figure}
\caption{The resistivity of the metallic phase as a function of $R_{n}$ in
Ga film experiment[5] (the diamonds are the datapoints from fig.2 of 
this reference) and the best fit of eqn.(25) (the dashed curve). (We have not
displayed those points of fig.2 in Ref.5 which are in the 
insulating regime and, expectedly, deviate from the best fit shown here.)}
\label{fig2}
\end{figure}


\begin{thebibliography}{99}
\bibitem{n1} One may equivalently start from the action in eqn.(10) as well.
\bibitem{n2} For this part of the problem, we work in the gauge $A_{0}=0,
\nabla_{\alpha}A^{\alpha}_{i}=0$. Choice of a particular gauge 
shifts the phase boundary by factors of order unity.
\bibitem{n3} Ref.\cite{it} and references cited therein.
\bibitem{n4} This differs from the usual $ln5$ because of our particular
choice of the gauge.
\bibitem{n5} For this part, we have replaced $ln(\frac{2}{J\Delta\tau})$
by $\frac{1}{2J\Delta\tau}$ for convenience. These two factors really 
come about at different limits of the magnitude of $J\Delta\tau$ 
while performing the Villain transformation. Please see Ref.\cite{bd} for more 
details. We believe that this replacement does not create any major problem,
since $\Delta\tau$ simply sets the energy scale here.
\bibitem{n6} A strong possibility is that the fluctuations
will suppress the critical value $\alpha_{cr}$ discussed in the context of
eqn.(19) and thus stabilise the BM phase even further.
\end{thebibliography}
\end{document}